\documentclass[useAMS,usenatbib,usegraphicx,letterpaper]{mn2e}
\usepackage{times,amssymb,hyperref,subfigure,epsfig,aas_macros}

\usepackage[totalwidth=480pt,totalheight=680pt,layoutvoffset=0.5cm]{geometry}

\begin{document}

\title[Warm exo-Zodi from cool exo-Kuiper belts]{Warm exo-Zodi from cool exo-Kuiper
  belts: the significance of P-R drag and the inference of intervening planets}

\author[Grant M. Kennedy \& Anjali Piette]{Grant M. Kennedy\thanks{Email:
    \href{mailto:gkennedy@ast.cam.ac.uk}{gkennedy@ast.cam.ac.uk}}$^1$, Anjali Piette$^2$ \\
  $^1$ Institute of Astronomy, University of Cambridge, Madingley Road, Cambridge CB3 0HA, UK \\
  $^2$ Pembroke College, Trumpington Street, Cambridge CB2 1RF, UK \\
}
\maketitle

\begin{abstract}
  Poynting-Robertson drag has been considered an ineffective mechanism for delivering
  dust to regions interior to the cool Kuiper belt analogues seen around other Sun-like
  stars. This conclusion is however based on the very large contrast in dust optical
  depth between the parent belt and the interior regions that results from the dominance
  of collisions over drag in systems with detectable cool belts. Here, we show that the
  levels of habitable zone dust arising from detectable Kuiper belt analogues can be tens
  to a few hundreds of times greater than the optical depth in the Solar Zodiacal
  cloud. Dust enhancements of more than a few tens of `zodi' are expected to hinder
  future Earth-imaging missions, but relatively few undetectable Kuiper belts result in
  such levels, particularly around stars older than a few Gyr. Thus, current mid to
  far-IR photometric surveys have already identified most of the 20-25\% of nearby stars
  where P-R drag from outer belts could seriously impact Earth-imaging. The LBTI should
  easily detect such warm dust around many nearby stars with outer belts, and will
  provide insight into currently unclear details of the competition between P-R drag and
  collisions. Given sufficient confidence in future models, the inevitability of P-R drag
  means that the non-detection of warm dust where detectable levels were expected could
  be used to infer additional dust removal process, the most likely being the presence of
  intervening planets.
\end{abstract}


\section{Introduction}\label{s:intro}

As the regions deemed most likely to harbour alien life, the habitable zones of other
stars are becoming of increasing interest. Major efforts to find potentially habitable
planets are underway \citep[e.g. \emph{Kepler},
PLATO,][]{2003SPIE.4854..129B,2014ExA....38..249R}, with the goal of pushing the
detection limits towards true Earth-analogues. Coupled to this interest has been parallel
work on exo-Zodiacal dust, both as a potentially useful dynamical tracer that reveals
unseen planets \citep[e.g.][]{2008ApJ...686..637S}, and as a source of noise and
confusion that may hinder the direct detection and characterisation of such planets
\citep{2010A&A...509A...9D,2012PASP..124..799R,2014ApJ...795..122S,2015ApJ...799...87B}.

Habitable zone (HZ) dust is indeed seen around other stars
\citep[e.g.][]{2005Natur.436..363S,2010ApJ...714L.152F}, though these detections are
limited by the photometric methods used to relatively high dust levels that are rare
\citep{2013MNRAS.433.2334K}. In general the origin of this exo-Zodiacal (`exo-zodi') dust
is unknown. In some cases, particularly for main-sequence stars, the levels are
sufficiently extreme that the dust cannot simply originate in massive analogues of our
Asteroid belt because collisional decay would have ground the dust levels well below
those observed over the stellar lifetime \citep{2007ApJ...658..569W}. Possible solutions
are that the dust was created in a recent collision \citep[i.e. is transient,
e.g.][]{2005Natur.436..363S,2012ApJ...751L..17M}, or that the dust is being continuously
delivered in the form of comets from more distant regions
\citep[e.g.][]{2007ApJ...658..569W,2012A&A...548A.104B}.

Curiously, the mechanism responsible for delivering much of the Zodiacal dust from
exterior regions to the Earth's vicinity, Poynting-Robertson (P-R) drag, can be ruled out
in almost all extra-Solar cases. The reason is that particles spiralling in by P-R drag
are also subject to collisional destruction and subsequent radiation-pressure blowout
\citep[for stars more luminous than the Sun, see][]{2011A&A...527A..57R}. The fate of
dust created in a parent belt of asteroids or comets (i.e. a `debris disk') therefore
depends on the relative importance of P-R drag and collisions. This issue was explored by
\citet{2005A&A...433.1007W}, who used a simple model of P-R drag for dust of a single
particle size \citep[see also][]{1999ApJ...527..918W}. The key points were: i) the denser
the parent belt, the greater the contrast in optical depth between the parent belt and
the inner regions, ii) the denser the parent belt, the greater the absolute dust level in
the inner regions up to some maximum level, and iii) the debris disks that are detected
around other stars are generally collision dominated, rather than P-R drag dominated, and
hence P-R drag could be considered insignificant. The motivation was to show that the
invocation of extra forces (e.g. unseen planets) was not necessary to explain why the
regions interior to some cool disks were not filled in. That is, the model was largely
considered within the context of the relative levels of dust in the parent belts and in
the interior regions.

An aspect of the model that was not explored in detail by \citet{2005A&A...433.1007W} was
the absolute level of interior dust, in part because this was not the primary motivation,
but also because observations at the time were not sensitive enough to detect the dust
levels predicted. However, new instruments are pushing the sensitivity limits for warm
dust to levels low enough that P-R drag as an origin of habitable zone dust is becoming
relevant. While the photometric methods used to discover bright dust are limited to
mid-IR disk to star flux ratios greater than about 10\% (at 3$\sigma$), the Keck
Interferometer Nuller \citep[KIN,][]{2012ApJ...748...55S} was able to reach levels closer
to 1\% \citep{2011ApJ...734...67M,2014ApJ...797..119M}, and the Large Binocular Telescope
Interferometer \citep[LBTI,][]{2009AIPC.1158..313H,2015ApJ...799...42D} is expected to go
below 0.1\%.

Specifically, \citet{2014ApJ...797..119M} report new results from the KIN, showing that
the level of warm dust seen around nearby stars tends to be higher in systems with
detections of cool Kuiper belt analogues. The surprising aspect is that in most cases the
levels tend to be similar, a few hundred times the Solar System level. The typical
explanation for a direct link between warm and cool belts is comet delivery by planet
scattering \citep[e.g.][]{2012MNRAS.420.2990B,2012A&A...548A.104B}, but the outcome of
this scenario depends sensitively on various parameters such as the planetary system
architecture, so is an unlikely explanation. In contrast, the competition between P-R
drag and collisions naturally leads to warm dust levels that are relatively insensitive
to the properties of the parent belt, and \citet{2014ApJ...797..119M} show that their
results are consistent with levels expected from the \citet{2005A&A...433.1007W} model of
P-R drag delivery

Here, we consider the wider implications of these findings, including the results of
\citet{2014A&A...571A..51V} who developed a more sophisticated numerical model of the
competition between P-R drag and collisions. Their results are easily incorporated into
the analytic model of \citet{2005A&A...433.1007W}, which we use to make general
predictions of the warm dust levels in systems with Kuiper belt analogues.

\section{Model}\label{s:model}

\subsection{Optical depth interior to parent belt}\label{ss:tau}

The steady-state P-R drag model of \citet{2005A&A...433.1007W} calculates the dust
optical depth as a function of radius interior to a source region, a parent belt of
planetesimals where dust is created in collisions between larger bodies. It is an
analytic solution of the continuity equation, where the number of particles entering a
radial region equals that leaving plus that lost to collisions. The main assumptions were
that particles have a single size, and that all collisions were destructive and result in
permanent loss of the fragments due to stellar radiation pressure. The model does not
therefore apply to stars with insufficient luminosity or stellar wind pressure to blow
small dust out of the system \citep[see][for discussions of this
case]{2006A&A...455..987A,2011A&A...527A..57R,2012A&A...548A..86L,2014A&A...567A.127S}.

Particles at some outer radius $r_0$, representing the parent belt, have a face-on
geometric optical depth $\tau_0$, and the only remaining free parameters are the stellar
mass and the strength of the P-R drag on the particles. The latter is parameterised by
the ratio of the force felt by a particle due to radiation pressure to the gravitational
force, $\beta = F_{\rm rad}/F_\star \propto L_\star/M_\star$ (i.e. we assume that the
analogous stellar wind drag effect is relatively small). Small particles created on
initially circular orbits are blown out of the system on a dynamical timescale when
$\beta > 0.5$, so the model assumed $\beta = 0.5$.

A numerical version of this model was developed by \citet{2014A&A...571A..51V}, who
included a dust size distribution and accounted for the fact that grains of different
sizes are affected to differing degrees by both radiation pressure and P-R drag, and also
have different collisional lifetimes due to differences in orbital eccentricity, strength
and their relative numbers. Compared to the analytic model, the only significant
difference was that the dust levels in the regions interior to the parent belt were
nearly an order of magnitude lower. The reason is that the smallest bound grains are
created on highly eccentric orbits meaning that collisions are more frequent and at
higher relative velocities, so a given particle can be destroyed by a smaller (more
common) impactor. They found that interior to the parent belt, the size distribution is
dominated by grains near the blowout size, which is consistent with the single grain size
assumed in the analytic model. They only explored cases in which the parent belt was
relatively massive, so it seems likely that this conclusion arises due to the dominance
of collisions immediately interior to the parent belt, where larger grains are heavily
depleted by the more numerous near-blowout size grains. In the case where collisions are
much less important larger grains would be expected to dominate because of their slower
P-R drag timescale. Our focus here is to relate observations to detectable parent belts,
so the approximation of a single grain size at the blowout limit is reasonable, but more
numerical work is needed to see how the results change with the parent belt optical
depth.

For relatively dense parent belts a simple way to include the effect of elliptical orbits
in the analytic model is to change the collision timescale by a multiplicative factor,
$k$. The solution to the continuity equation for the optical depth, $\tau(r)$, is
\begin{equation}\label{eq:tau}
  \tau(r) = \tau_0 / \left[ 1+ 4 \eta_0 \left( 1 - \sqrt{r/r_0} \right) \right] ,
\end{equation}
\begin{equation}\label{eq:eta}
  \eta_0 = 5000 \tau_0 \sqrt{ (r_0/a_\oplus) ( M_\odot/ M_\star ) } /(\beta k) .
\end{equation}
These equations only differ from those in \citet{2005A&A...433.1007W} by the factor
$k$. Given the results of \citet{2014A&A...571A..51V}, we can assume that $\beta = 0.5$
because these grains dominate the size distribution. We set $k=1$ to represent the
original \citet{2005A&A...433.1007W} model, which is consistent with the KIN results and
is referred to as the ``low collision rate'', or set $k = 1/7$ to match the numerical
model of \citet{2014A&A...571A..51V}, which we refer to as the ``high collision
rate''. In the absence of the KIN results we would prefer the high collision rate because
it was derived by a model thought to be more realistic. However, an apparently lower
collision rate could arise because the numerical model does not include all important
physical effects, for example if not all collisions result in destruction and subsequent
blowout \citep[e.g.][]{2014A&A...566L...2K}. One of our goals is therefore to make
predictions for each case so that observations can empirically find which is more
realistic.

To convert the dust optical depth given by eq. (\ref{eq:tau}) into an observable flux
density, we follow \citet{2005A&A...433.1007W} and assume the emission is from grains
that behave as black bodies. Though the dust can in some cases be small enough to be
hotter than a black body at a given distance from the star, tests using realistic grain
properties find that this approximation is reasonable (and if anything an underestimate
of the flux) when calculating mid-IR emission; for A-type stars the smallest grains are
large enough to behave approximately as blackbodies, and for later type stars the
inefficient emission at wavelengths longer than the grain size is offset by their
increased temperatures and consequently greater fluxes.

\subsection{Observed Kuiper belt analogues}

To connect observations of Kuiper belt analogues to the parent belt properties in the
model, we must convert them to our model parameters $r_0$ and $\tau_0$. The radius can be
easily derived from the dust temperature of an observed disk assuming blackbody
properties, with the caveat that these estimates are typically a factor of a few too low,
depending on the spectral type of the star
\citep[e.g.][]{2012ApJ...745..147R,2013MNRAS.428.1263B,2014ApJ...792...65P}. In any case,
as we show below, an uncertain radius has a relatively small impact on the model
predictions. The optical depth can be estimated from the fractional luminosity $f=L_{\rm
  disk}/L_\star=\sigma/(4 \pi r_0^2)$, where $\sigma$ is the surface area of emitting
dust, with the assumption of some fractional disk width via $\sigma = 2 \pi r_0 \Delta r
\tau_0$. Again, as we show below the habitable zone dust levels are relatively
insensitive to the optical depth of the parent belt, so here we assume $\Delta r = 0.5
r_0$, and thus $\tau_0 = 4 f$.

For a sample of known Kuiper belt analogues we use stars from the Unbiased Nearby Stars
(UNS) sample \citep{2010MNRAS.403.1089P}. Most of these have been observed with
\emph{Spitzer} and/or \emph{Herschel}
\citep[e.g.][]{2005ApJ...620.1010R,2008ApJ...674.1086T,2013A&A...555A..11E,2014MNRAS.445.2558T},
and here we use disk properties derived from this far-IR photometry to show a
representative disk population. The stellar photospheres have been modelled with PHOENIX
AMES-Cond models \citep{2005ESASP.576..565B} and the excess fluxes with blackbodies to
derive the dust temperatures and luminosities, which are the quantities of interest here
\citep[e.g.][]{2012MNRAS.421.2264K,2012MNRAS.426.2115K,2014MNRAS.445.2558T}. These belts
have radii of a few au to a few hundred au, and fractional luminosities between $10^{-6}$
and $10^{-3}$.

\section{Habitable zone dust levels}\label{s:hz}

\subsection{An example}\label{ss:hzeg}

% figs/hztau.pro
\begin{figure}
  \begin{center}
    \hspace{-0.15cm} \includegraphics[width=0.48\textwidth]{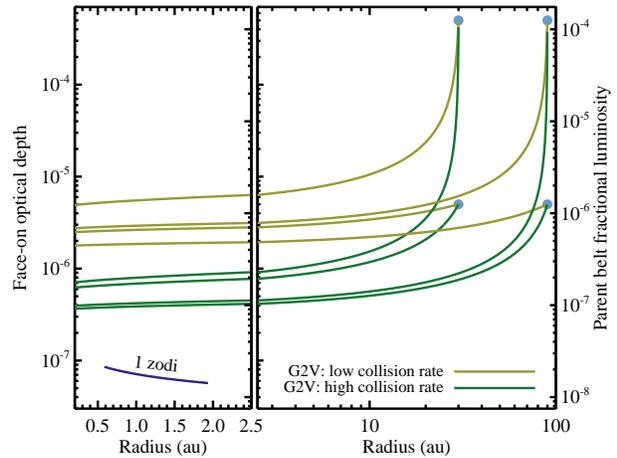}
    \caption{Examples of radial optical depth profiles interior to parent belts around a
      Sun-like star. The left-hand panel shows a linear radial scale while the right-hand
      panel is logarithmic. Four different parent belts are shown as dots, with $r_0=30$
      and 90 au, and $\tau_0=5 \times 10^{-6}$ and $5 \times 10^{-4}$. The darker 
      lines show the high collision rate ($k=1/7$) and lighter lines show the low
      collision rate ($k=1$). The Solar System's Zodiacal cloud level (`1 zodi') is also
      shown from 0.5 to 2 au. The dust optical depth in the habitable zone is
      near-constant, and depends strongly on the collision rate and weakly on the optical
      depth in the parent belt.}\label{fig:tau}
  \end{center}
\end{figure}

Fig. \ref{fig:tau} shows radial optical depth profiles for a Sun-like star for both
collision rates and several different parent belt parameters. The outer belt parameters
are chosen to represent a relatively bright Kuiper belt analogue, and one near the
current detection limit. The optical depth of the Solar Zodiacal cloud (i.e. 1 `zodi'
unit of dust) is shown from 0.5 to 2 au \citep{1998ApJ...508...44K}.

Dust is created in the parent belt with optical depth $\tau_0$, and then moves towards
the star due to P-R drag, being depleted by collisions on the way. The collision rate
depends on the optical depth, so eventually P-R drag dominates. The P-R drag time is
$\propto 1/r$ so the optical depth approaches a constant level with radius. Thus, the
optical depth interior to very tenuous parent belts (where collisions are unimportant) is
also constant.

For denser parent belts the interior levels become insensitive to the parameters. More
dust drifts inwards from denser belts, but this extra dust is also collisionally depleted
more rapidly. For fixed stellar parameters and parent belt radius, the interior optical
depth reaches a maximum optical depth near the star as $\tau_0$ is increased
\citep[i.e. $n_0 \gg 1$ and $\tau (r)$ is independent of
$\tau_0$,][]{2005A&A...433.1007W}. This insensitivity to parent belt properties is an
important aspect of the model that is not expected in other scenarios, where stochastic
behaviour and/or strong dependence on system architecture are expected to lead to a wide
variety of habitable zone dust levels
\citep[e.g.][]{2013MNRAS.433.2334K,2012MNRAS.420.2990B}. However, the inner dust levels
depend strongly on the collision rate assumed (i.e. $k$). For the examples in
Fig. \ref{fig:tau} the high collision rate predicts dust levels that are approximately 10
times the Solar Zodiacal level, and the low collision rate levels about 50 times
larger. In either case, in the absence of effects that remove the inward moving dust, the
regions interior to Kuiper belt analogues seen around other stars (with $f>10^{-6}$) will
have dust levels that are significantly higher than the Solar Zodiacal level.

In the outer Solar System the dust level interior to the Edgeworth-Kuiper belt, with
$\tau \sim 10^{-7}$ \citep{2012A&A...540A..30V}, is thought to be constant in to about 10
au, where most of the particles are ejected by Saturn and Jupiter \citep[estimates of the
fraction reaching Earth's vicinity depend on grain size, and vary from
0-20\%,][]{1996Icar..124..429L,2010AJ....140.1007K}. The optical depth of the Asteroid
belt is similar \citep{2002aste.conf..423D}, but most of the dust that spirals in towards
the Earth is not lost. Given the uncertainties in their inferred optical depths and
models, both have been considered a source for our Zodiacal cloud
\citep{1996Icar..124..429L,2001Icar..152..251G}. In reality, both contribute at some
level, and material from Jupiter-family comets serves as a third and equally important
source \citep[i.e. dynamical rather then P-R drag delivery from the Kuiper
belt,][]{1967SAOSR.239.....W,2010ApJ...713..816N}. Thus, external observations of the
Solar System's dust complement (that might not detect the gap between $\sim$3-5 au),
might reasonably conclude that the Zodiacal dust originates in the Kuiper belt and is
delivered by P-R drag. However, our knowledge of the actual architecture shows that
planets can strongly influence the habitable zone dust level, both depleting it relative
to simple expectations by removing dust as it spirals in, and enhancing it by allowing
other delivery mechanisms.

\subsection{Model predictions}

% figs/hztau.pro
\begin{figure*}
  \begin{center}
   \hspace{-0.015cm} \includegraphics[width=0.48\textwidth]{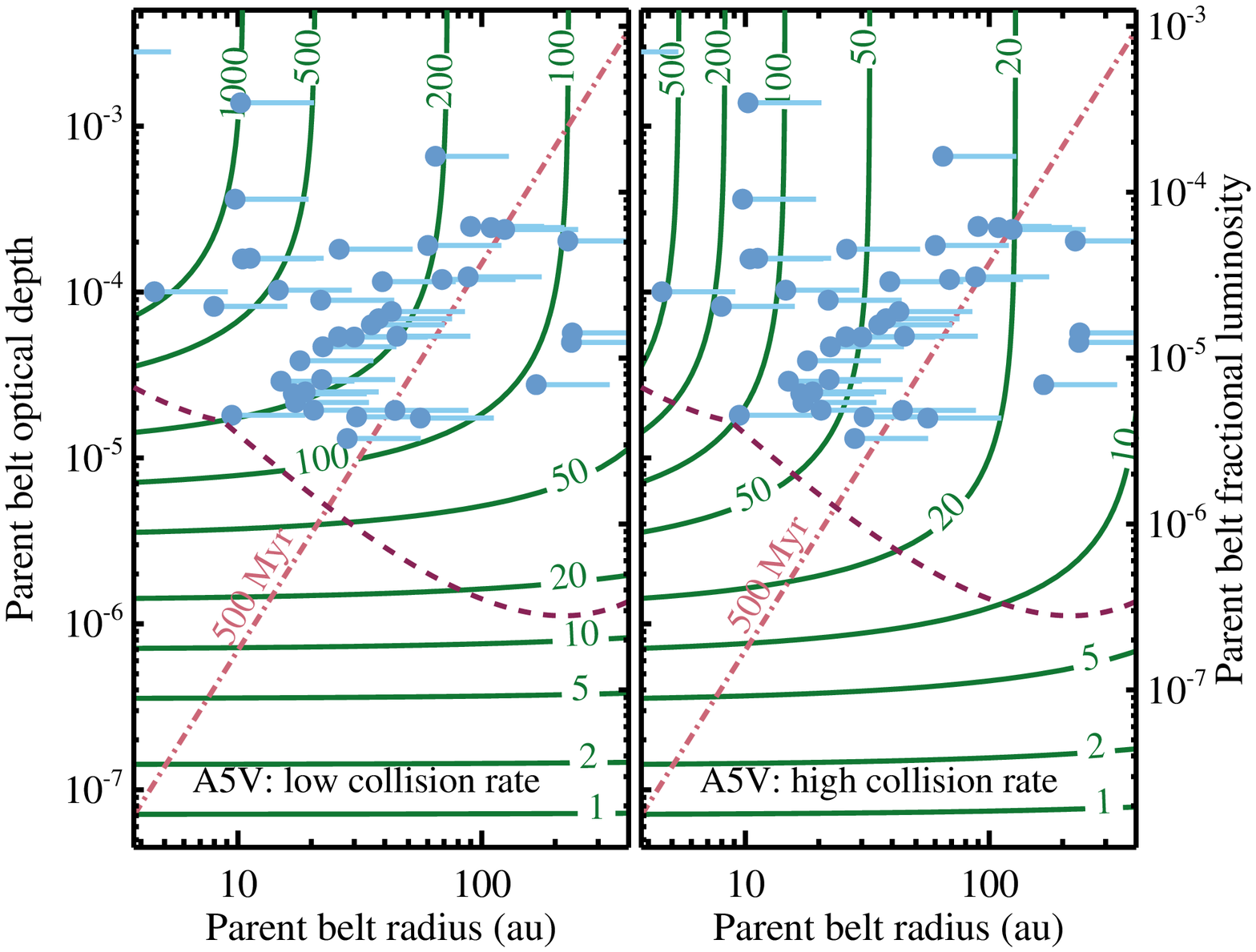}
    \hspace{0.4cm} \includegraphics[width=0.48\textwidth]{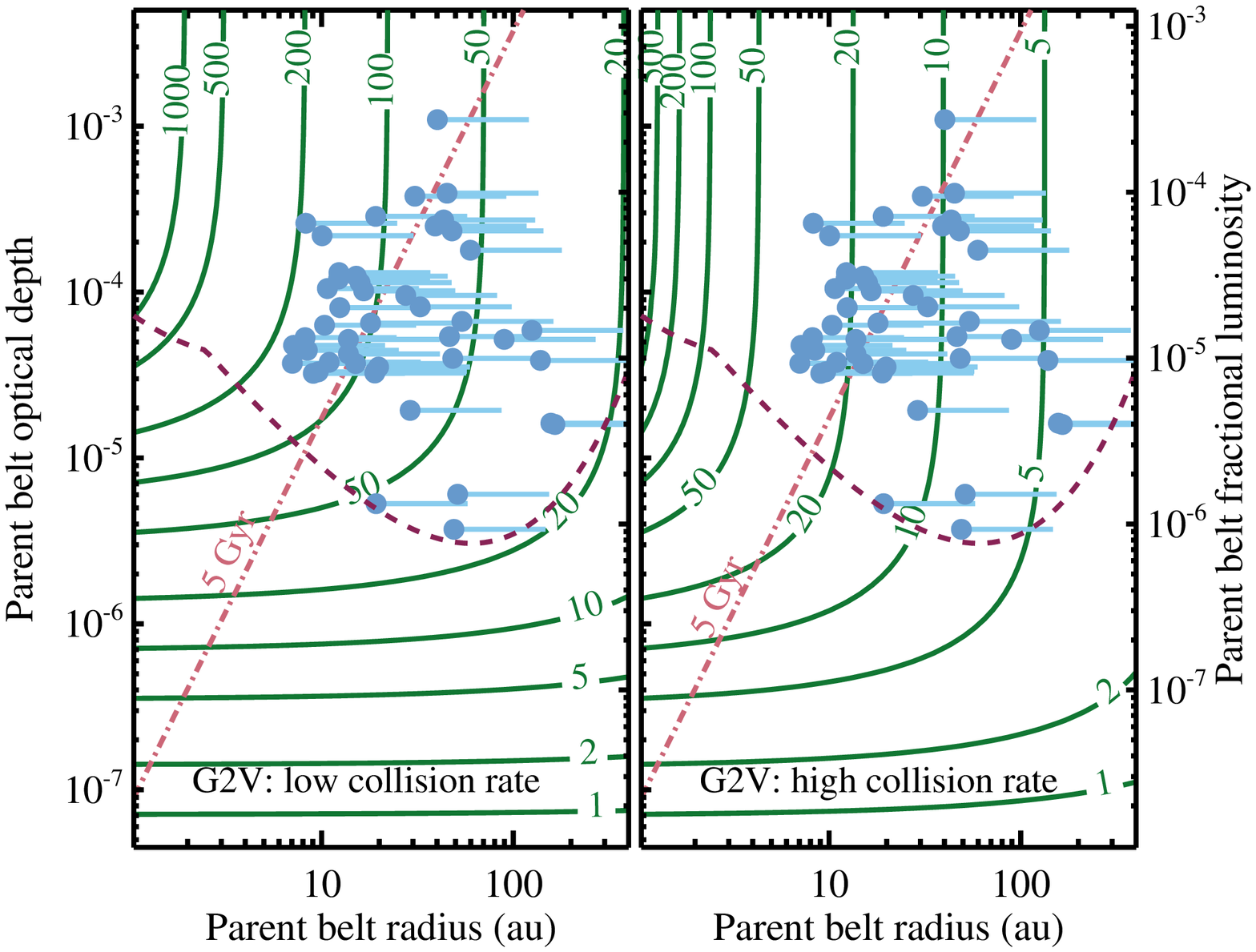}
    \caption{Dust levels at the EEID for an A-type (3.7 au) and a Sun-like (1 au) star
      for a range of $r_0$ and $\tau_0$. Contours show the optical depth relative to the
      Solar System level at 1 au of $7.12 \times 10^{-8}$. Dots mark known Kuiper belt
      analogues around nearby stars and connected lines show the probable increase in
      size due to non-blackbody emission from small dust. An approximate photometric
      detection limit is shown by the dashed line. Dot dashed lines show the maximum
      fractional luminosity as a function of radius for stellar ages of 500 Myr and 5 Gyr
      (as labelled). For each panel, the left sub-panel shows results for the low
      collision rate ($k=1$), and the right sub-panel shows results for the high
      collision rate ($k=1/7$). The left end of the radial scale stops at the EEID in
      each case.}\label{fig:hztaug}
  \end{center}
\end{figure*}

Fig. \ref{fig:hztaug} shows predicted dust levels in the habitable zone for an A-type and
a Sun-like star. We use the concept of the ``Earth-equivalent insolation distance''
(EEID) as a simple definition of the habitable zone location, which is simply equal to
$\sqrt{L_\star/L_\odot}$ in units of au \citep[i.e. 1 au for the
Sun,][]{2012PASP..124..799R}. In each panel contours show the dust level at the EEID,
relative to the Solar System optical depth at 1 au of $7.12 \times 10^{-8}$
\citep{1998ApJ...508...44K}, for a range of parent belt optical depths and radii. That
is, at the EEID (but not necessarily elsewhere) these contours are the same as the
``zodi'' units defined in \citet{2015ApJS..216...23K} for the purposes of LBTI
modelling. Sub-panels for both the high and low collision rate models are shown. The
symbols mark known Kuiper belt analogues seen around nearby stars from the sample of
\citep{2010MNRAS.403.1089P}, including a `tail' that reflects the approximate systematic
uncertainty in the dust radial location. Vertical uncertainties are relatively small
(roughly a factor $\pm$2) as any model that fits the disk photometry will have a similar
integrated luminosity. A representative detection limit for the parent belts around
nearby stars with mid and far-IR photometry (24 and 100 $\mu$m) is shown \citep[computed
as in][]{2008ARA&A..46..339W}, where detections are possible above the dashed line, and
differs for A and G-type stars because their luminosities are different. At low parent
belt densities the dust level contours are roughly constant with radius because P-R drag
dominates. At high densities collisions dominate and the habitable zone dust level
depends only on the radial proximity of the parent belt to the habitable zone.

Thus, P-R drag of dust from known Kuiper belt analogues can result in significant HZ dust
levels relative to the Solar System level, regardless of the assumed collision rate. The
results clearly depend on the adopted collision rate; the low collision rate implied by
the KIN results suggests that the HZ dust level is typically 50-500 times the Solar
Zodiacal level for detectable parent belts around Sun-like and A-type stars. The higher
collision rate results in typical levels closer to 10-100 times the Solar Zodiacal level.

The nature of the photometric detection limits means that there is a region of the
parameter space in all cases where HZ dust levels in excess of a few tens of times the
Solar System level can arise from belts that cannot be detected by photometry. These
parent belts tend to be close to the habitable zone however, and therefore decay by
collisions faster than belts that are farther out, mostly due to higher relative
velocities. The approximate maximum optical depth of a parent belt in collisional
equilibrium is shown in each panel, for average stellar ages of 500 Myr and 5 Gyr, using
planetesimal properties that fit their observed brightness evolution
\citep{2007ApJ...663..365W,2011MNRAS.414.2486K}. The vertical position of these lines
varies as $1/t_{\rm age}$, and for a given age few belts are seen or expected above such
lines \citep[e.g.][]{2007ApJ...658..569W}. Thus, the space where belts are allowed by
collisional evolution but are not detectable and result in significant HZ dust levels is
relatively small, and decreases in size as the age of the star increases.

\section{Discussion}\label{s:disc}

Poynting-Robertson drag is one of many ways to produce habitable zone dust, the main
alternatives being material produced locally in collisions or dynamical comet delivery by
planet scattering. Thus, where a Kuiper belt analogue is known the exo-zodi levels
discussed above are maximum P-R drag induced dust levels that can be compared with
observations to draw conclusions about the dust origin. That is, dust levels
  significantly greater than predicted require a different origin.

  For example, the most extreme system known to harbour warm and cool dust populations is
  the F2V star $\eta$ Corvi
  \citep[e.g.][]{2005ApJ...620..492W,2009A&A...503..265S,2015ApJ...799...42D}, where the
  location and optical depth of the warm belt is roughly 1 au and $10^{-3}$, and 150 au
  and $10^{-4}$ for the cool belt. Fig. \ref{fig:tau} shows that the HZ optical depth
  expected from any parent belt from P-R drag is lower than $10^{-5}$, and therefore that
  the warm dust is not delivered from the outer belt by P-R drag. Indeed, the only
  systems where such bright warm dust may originate in cool belts are those with low
  stellar luminosities, where collisions during inspiral do not result in removal, and
  thus the interior optical depth remains approximately the same as in the parent belt
  \citep[e.g. $\epsilon$ Eri,][]{2011A&A...527A..57R}. In general therefore, detections
  of warm dust well above the predictions of the P-R drag models around Solar and
  earlier-type stars require a different origin. In the case of $\eta$ Crv the favoured
  scenario is that comets are being scattered inward from the cool belt by unseen planets
  \citep[e.g.][]{2007ApJ...658..569W,2012ApJ...747...93L}.

\subsection{Habitable zone dust levels}

It seems probable that all stars host cool outer debris disks, with only the brightest
20-25\% being detectable with current methods
\citep[e.g.][]{2008ApJ...674.1086T,2006ApJ...653..675S,2013A&A...555A..11E,2014ApJ...785...33S,2014MNRAS.445.2558T}. Indeed,
this was the assumption made in various population models that reasonably reproduce the
statistics of such disks
\citep[e.g.][]{2007ApJ...663..365W,2011MNRAS.414.2486K,2013ApJ...768...25G}. Given the
inevitability of P-R drag, this population of cool outer belts implies the existence of a
corresponding population of warm dust around all stars. Combined with the results in
Fig. \ref{fig:hztaug}, the 20-25\% of stars with detected cool belts can therefore be
used to make some useful statements about HZ dust levels.

For A-type stars, Fig. \ref{fig:hztaug} shows that detectable cool belts result in HZ
dust levels at least 20-100 times greater than in the Solar System (i.e. most known outer
belts are above these contours). Thus, among the remaining 75-80\% of stars the HZ dust
levels due to P-R drag are below these levels. For G-type stars the predicted HZ dust
levels are lower, and the 75-80\% or so of Sun-like stars with no known outer belts have
HZ dust below 10-50 times the Solar System level.

Various studies have estimated the impact of HZ dust on future missions that will attempt
to detect and characterise Earth-like planets around other stars. The two most recent
studies seem to reach a consensus that, with the appropriate survey strategy and under
the assumption of smooth exo-zodi, this impact is a relatively weak function of dust
brightness, and that for levels below a few tens of times the Solar System level a
mission will not be seriously hindered
\citep{2014ApJ...795..122S,2015ApJ...799...87B}. The possible impact of non-axisymmetric
structure is less certain however, since inhomogeneities in exo-Zodi surface brightness
may themselves be mistaken for planets
\citep[e.g.][]{2004ApOpt..43.6100L,2010A&A...509A...9D}. Such structures may of course
also be signposts of the dynamical influence of unseen planets
\citep[e.g.][]{2008ApJ...686..637S}. Whether such structure presents a major barrier
depends both on the typical level of non-axisymmetry and the specifics of the
Earth-imaging mission, but could result in a tolerable exo-Zodi level closer to ten times
the Solar System level.

Based on our HZ dust predictions, Earth-imaging efforts for Sun-like stars are unlikely
to be seriously affected by dust coming in from undetected outer belts, but most of the
20-25\% of systems with detected outer belts would be considered unsuitable targets for
this reason. For A-type stars such a statement is less certain, as it is possible that
10-30 au belts just below the level of detectability (whose frequency is unknown), will
contribute to HZ dust levels. Collisional depletion of the parent belts over time means
that this issue can be mitigated to some degree by avoiding young stars. It is unlikely
that such belts make up a large fraction of the population however, so the fraction of
unsuitable systems is probably not significantly greater than 25\%.

Thus, while a non-negligible fraction of stars are expected to be poorly suited to
Earth-imaging/characterisation due to P-R drag from outer belts, the detection limits of
extant far-IR surveys means that among stars within a few tens of parsec most of these
systems are already known. As noted above however, P-R drag is by no means the only
origin of habitable zone dust. Thus, in the absence of other effects (see next section)
the levels predicted by detections of cool outer belts represent a lower limit on dust in
a given system.

\subsection{Inference of unseen planets}

The inward transport of dust from a known parent belt by P-R drag is inevitable, so
additional processes must also be invoked if HZ dust levels could be shown to be
significantly lower than expected. That is, something may be needed to remove the dust
somewhere between the parent belt and where detection was attempted. By analogy with the
Solar System, the most obvious reason is of course accretion or ejection of that dust by
intervening planets
\citep{1996Icar..124..429L,1999AJ....118..580L,1999ApJ...527..918W,2012A&A...540A..30V}.
Whether the dust is accreted or ejected will to first order depend on the ratio of the
escape velocity from the planet to the local escape velocity from the star. Thus, a
further implication is that close-in planets (e.g. those that transit their stars) are
also those most likely to accrete dust, rather than eject it, perhaps with observable
effects. The limits set by the KIN were unfortunately not stringent enough to require the
explanation of dust removal by other processes for any systems, their Fig. 10 shows that
the non-detections are close to or above the levels predicted by the low collision rate
model.

Perhaps the most promising system where such a test could be made is around the
``retired'' A-type star $\kappa$ CrB. Though the spectral type is K0, the star is a
sub-giant and for our purposes can be treated as an A-type star because stellar
luminosity and mass are the important model parameters here. This system has a bright
outer dust belt that extends from 20-40 au out to 165-220 au depending on the disk model
\citep{2013MNRAS.431.3025B}, and at least two companions. One at 2.7 au has been well
characterised by radial velocity \citep{2008ApJ...675..784J}, and another is inferred to
exist at greater distance due to the existence of a residual acceleration in the radial
velocity data \citep{2013MNRAS.431.3025B}. With only a constant acceleration seen over 8
years, the outer companion lies beyond about 7 au, and therefore beyond the EEID of 3.5
au. This companion is inferred to be more massive than Jupiter, and resides at a radial
distance where the escape velocity from the star is lower, and can therefore easily eject
particles that interact with it. Thus, if this companion lies interior to the outer dust
belt it will eject much of the dust that would otherwise reach the habitable zone. Given
confidence in the P-R drag model, a sufficiently constraining non-detection of warm dust
around $\kappa$ CrB could therefore be used as further evidence of the second companion's
existence.

In the Solar System, the Asteroid belt and Jupiter family comets provide alternative
habitable zone dust sources, and it seems largely a coincidence that the Zodiacal cloud
has approximately the level expected for P-R drag from Kuiper belt dust. As noted in
section \ref{ss:hzeg}, an external observer with similar but moderately more sensitive
instruments than ours might conclude that the dust in the inner Solar System is
consistent with that expected from P-R drag from the Kuiper belt, and therefore that
invoking unseen planets is not necessary. Thus, while additional processes and dust
source regions may cause existing planets to be missed, they would not cause non-existent
planets to be inferred.

\subsection{Detectability with LBTI}\label{s:lbti}

An uncertainty in the above predictions for HZ dust levels is which of our two models is
more correct. That is, a significant step towards improving the model would be an
empirical calibration of the parameter $k$ across a range of stellar spectral types and
parent belt properties. We now show that these models will be strongly tested by the
LBTI, and hence aid further model development. The LBTI will survey a sample of $\sim$50
nearby stars, with a wide variety of spectral types, to look for habitable zone dust
\citep{2015ApJS..216...24W}. The sample includes stars known to host cool outer dust
belts; a few of these are Sun-like stars but most are earlier A-types.

\begin{figure*}
  \begin{center}
    \hspace{-0.25cm} \includegraphics[width=0.48\textwidth]{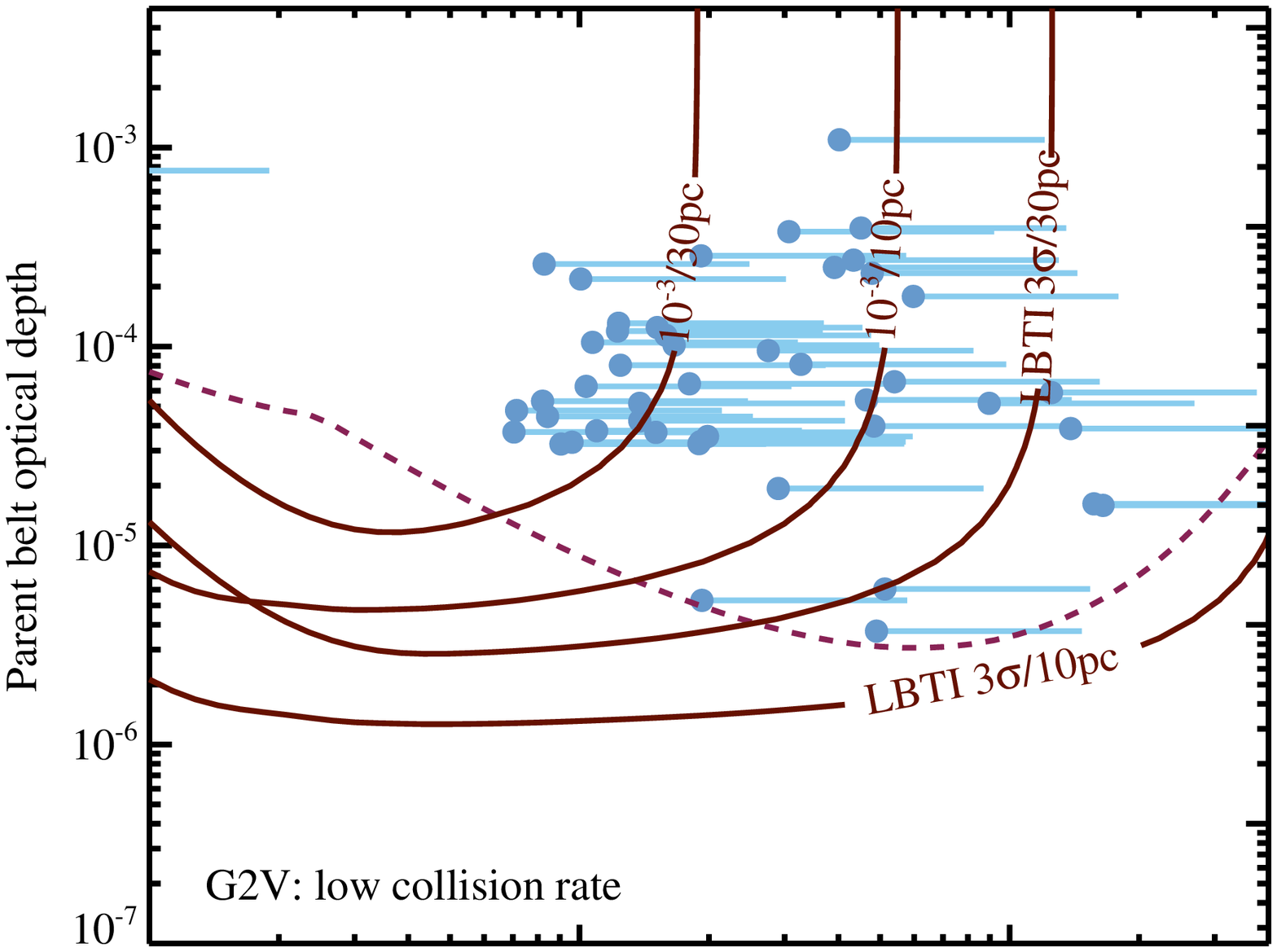}
    \hspace{-1.75cm} \includegraphics[width=0.48\textwidth]{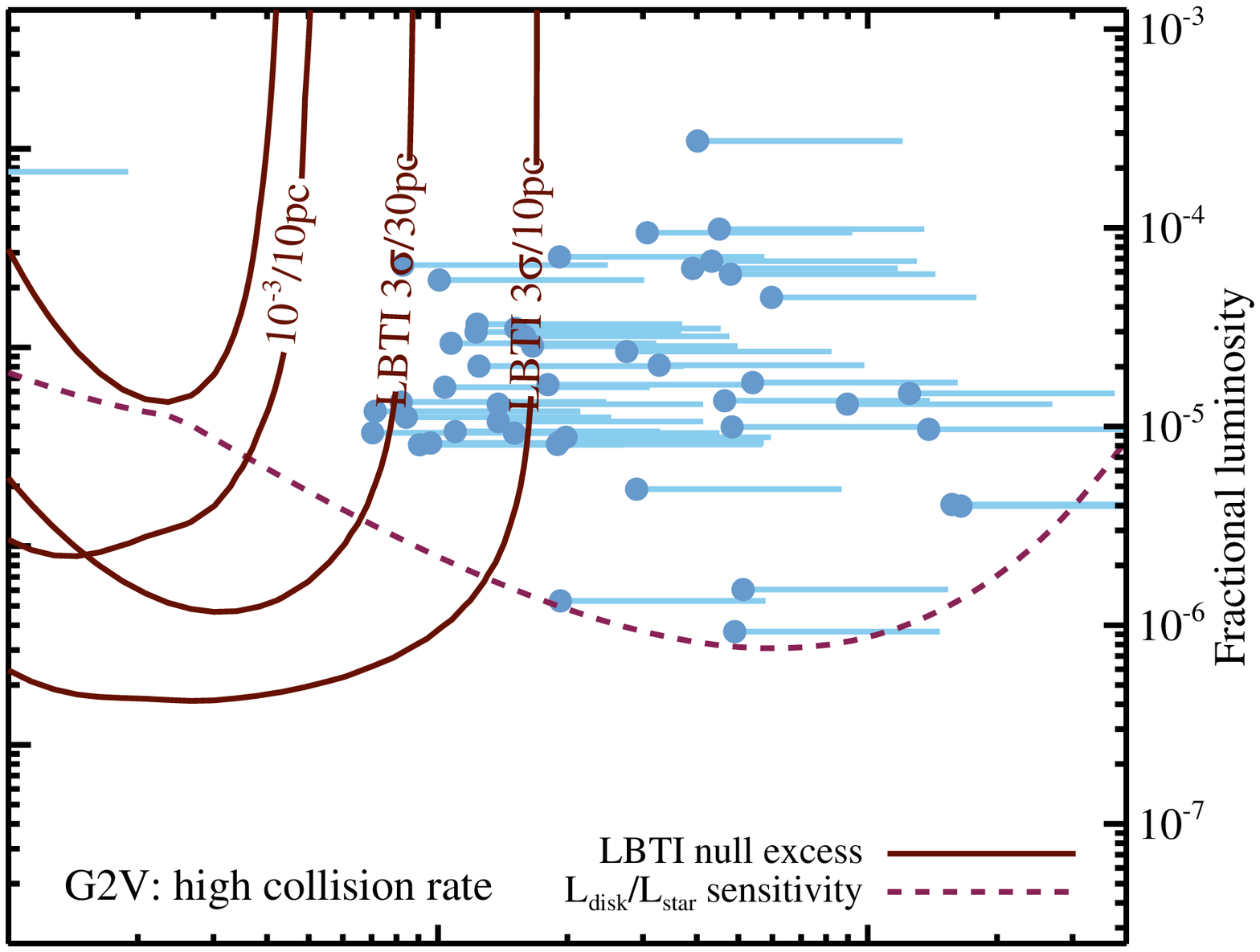} \\
    \vspace{-0.65cm}
    \hspace{-0.25cm} \includegraphics[width=0.48\textwidth]{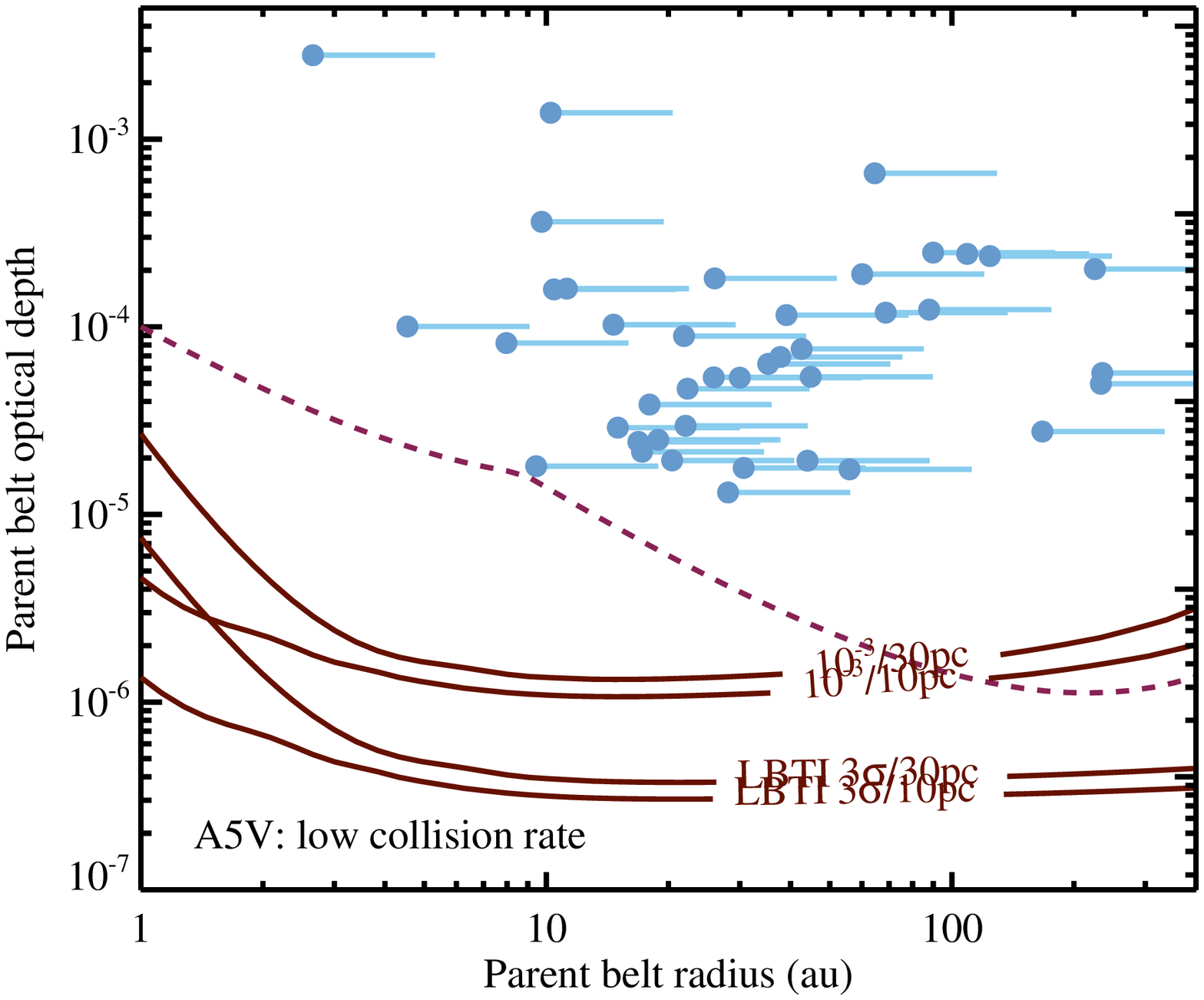}
    \hspace{-1.75cm} \includegraphics[width=0.48\textwidth]{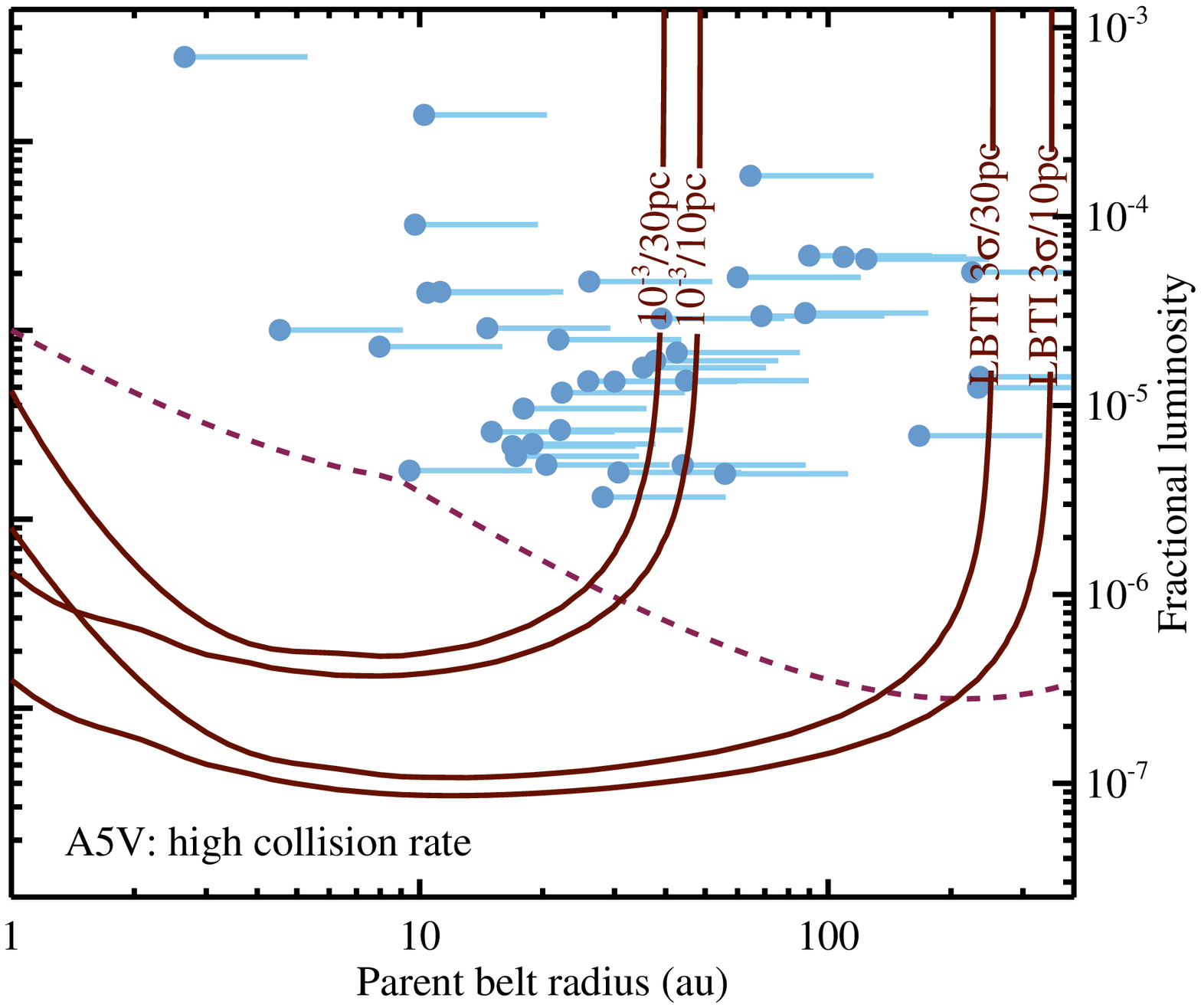}
    \caption{LBTI null excesses for a range of parent belt parameters, for two different
      sensitivities ($3 \times 10^{-4}$ and $10^{-3}$), and two stellar distances (10 and
      30pc, contours as labelled). The top row shows models for a Sun-like star, and the
      bottom row shows models for an A-type star. Left panels show models for the low
      collision rate, and right panels show models for the high collision rate. Dots mark
      known Kuiper belt analogues around nearby stars and connected lines show the
      probable increase in size due to non-blackbody emission from small dust. Dashed
      lines show approximate detection limits for parent belts around nearby stars. The
      null excesses always increase with parent belt optical depth, so for each distance
      the contours show the minimum optical depth for warm dust to be detectable at that
      level.}\label{fig:null}
  \end{center}
\end{figure*}

Briefly, the LBTI is a nulling interferometer that operates in the mid-IR (11 $\mu$m), so
interferes the light collected by each of the two LBT mirrors to attenuate the
starlight. Off-axis mid-IR emission at an angular scale of greater than $\sim$40 mas, for
example from warm dust, is transmitted. The attenuation of the starlight avoids the
problem of disentangling the dust emission from the stellar emission, which is the
limiting factor for discovery of habitable zone dust using purely photometric methods
(i.e. sets the dashed lines in Fig. \ref{fig:hztaug}). In practice, LBTI observations
involve various calibration steps, and for a description of these we refer the reader to
\citet{2015ApJ...799...42D}. The key point is that an LBTI measurement, known as the
``calibrated null depth'' or ``null excess'', is the ratio of the disk flux that is
transmitted though the LBTI fringe pattern to the star+disk flux. Thus, because the disk
is almost always much fainter than the star, the null excess can be thought of as
analogous to the disk to star flux ratio, but with the disk flux attenuated by the sky
projection of the LBTI transmission pattern
\citep{2011ApJ...734...67M,2015ApJ...799...42D}. We use the model outlined by
\citet{2015ApJS..216...23K}, and assume face-on disks to convert our model dust levels to
null excesses.

Null excesses are shown as contours in Fig. \ref{fig:null} again for a range of parent
belt properties, for both Sun-like and A-type stars and the low and high collision
rates. Again, known outer belts around nearby stars and photometric sensitivity limits
are also included. The projected LBTI 3$\sigma$ null excess sensitivity is
$3 \times 10^{-4}$, so these contours, and a higher level of $10^{-3}$, are shown. Because
the null excess depends on the angular scale of the disk, the contours show pairs at two
different distances (10 and 30 pc). The null excesses always increase with optical depth,
so for each distance the contours show the minimum parent belt optical depth for warm
dust to be detectable at that level.

Looking first at the low collision rate model (left panels), all of the known parent
belts lie in regions where the warm dust is detectable by the LBTI. In addition, for
A-stars there is a large swath of parameter space for LBTI detection of warm dust from
parent belts that are otherwise undetectable by photometry. A similar region exists for
Sun-like stars, but covers less parent belt parameter space. For this model warm dust is
therefore predicted to be detectable with LBTI, easily in most cases, for all systems
with known Kuiper belt analogues. Thus, if the KIN results of a direct connection between
warm and cool dust is the result of P-R drag and the low collision rate model applies
across AFG spectral types, the LBTI will strongly confirm this result.

In cases where no parent belt has been seen, warm dust originating in such a belt could
still be detected. The origin of this low level HZ dust will be ambiguous; mere detection
will be sufficiently difficult that further constraints on the radial extent, which could
provide clues to the origin, will be poor
\citep{2015ApJS..216...23K,2015ApJ...799...42D}.

For the high collision rate the results are very similar for A-type stars; nearly all
systems with known parent belts should result in LBTI detections. Detections of dust from
undetected parent belts also remains possible. For Sun-like stars however, only systems
with parent belts with radii $\lesssim$15 au should result in LBTI detections.

\section{Summary and Conclusions}\label{s:conc}

The development of new instruments means that P-R drag can no longer be considered an
insignificant effect for transporting dust to regions interior to the Kuiper belt
analogues seen around other stars. While the point that the contrast between a parent
belt and the inner regions is high remains, the absolute dust levels interior to known
parent belts can be hundreds of times higher than seen in the Solar Zodiacal cloud. We
have explored these levels using two flavours of an analytic P-R drag model, connecting
the properties of known Kuiper belt analogues to expected habitable zone dust levels.

Habitable zone dust levels greater than a few tens of times the Solar System level will
be detrimental to future missions to discover and characterise Earth-like planets around
other stars. Currently detectable Kuiper belt analogues are predicted to result in HZ
dust levels similar to this limit. Thus, mid and far-IR debris disk surveys have already
identified most of the 20-25\% of stars where P-R drag from outer belts could seriously
impact Earth-imaging. The existence of other processes that generate/deliver HZ dust of
course means that the 20-25\% is a lower limit.

A caveat to such predictions is that intervening planets may prevent this dust from
reaching the habitable zone, instead being ejected or accreted by intervening
planets. Thus, in systems with known outer belts the non-detection of such warm dust
could, given sufficient confidence in the models, be used to infer the existence of
unseen planets. Currently the models are uncertain however, so the goal of future work
should therefore be first to empirically calibrate the P-R drag models (e.g. with LBTI
observations) to aid model development, and then to interpret the observations within the
framework of such a model.

\section*{Acknowledgments}

We thank Rik van Lieshout and Mark Wyatt for useful discussions, Bertrand Mennesson for
sharing the KIN results ahead of publication, the reviewer for thoughtful and
constructive comments, and the LBTI Science and Instrument teams for providing some of
the motivation for this work. GMK is supported by the European Union through ERC grant
number 279973, and AP gratefully acknowledges support from an Undergraduate Research
Bursary from the Royal Astronomical Society.

%\bibliography{../ref} \bibliographystyle{apj}

\begin{thebibliography}{61}
\expandafter\ifx\csname natexlab\endcsname\relax\def\natexlab#1{#1}\fi

\bibitem[{{Augereau} \& {Beust}(2006)}]{2006A&A...455..987A}
{Augereau}, J.-C. \& {Beust}, H. 2006, \aap, 455, 987

\bibitem[{{Bonsor} {et~al.}(2012){Bonsor}, {Augereau}, \&
  {Th{\'e}bault}}]{2012A&A...548A.104B}
{Bonsor}, A., {Augereau}, J.-C., \& {Th{\'e}bault}, P. 2012, \aap, 548, A104

\bibitem[{{Bonsor} {et~al.}(2013){Bonsor}, {Kennedy}, {Crepp}, {Johnson},
  {Wyatt}, {Sibthorpe}, \& {Su}}]{2013MNRAS.431.3025B}
{Bonsor}, A., {Kennedy}, G.~M., {Crepp}, J.~R., {Johnson}, J.~A., {Wyatt},
  M.~C., {Sibthorpe}, B., \& {Su}, K.~Y.~L. 2013, \mnras, 431, 3025

\bibitem[{{Bonsor} \& {Wyatt}(2012)}]{2012MNRAS.420.2990B}
{Bonsor}, A. \& {Wyatt}, M.~C. 2012, \mnras, 420, 2990

\bibitem[{{Booth} {et~al.}(2013)}]{2013MNRAS.428.1263B}
{Booth}, M. {et~al.} 2013, \mnras, 428, 1263

\bibitem[{{Borucki} {et~al.}(2003){Borucki}, {Koch}, {Lissauer}, {Basri},
  {Caldwell}, {Cochran}, {Dunham}, {Geary}, {Latham}, {Gilliland}, {Caldwell},
  {Jenkins}, \& {Kondo}}]{2003SPIE.4854..129B}
{Borucki}, W.~J., {Koch}, D.~G., {Lissauer}, J.~J., {Basri}, G.~B., {Caldwell},
  J.~F., {Cochran}, W.~D., {Dunham}, E.~W., {Geary}, J.~C., {Latham}, D.~W.,
  {Gilliland}, R.~L., {Caldwell}, D.~A., {Jenkins}, J.~M., \& {Kondo}, Y. 2003,
  in Society of Photo-Optical Instrumentation Engineers (SPIE) Conference
  Series, Vol. 4854, Society of Photo-Optical Instrumentation Engineers (SPIE)
  Conference Series, ed. J.~C. {Blades} \& O.~H.~W. {Siegmund}, 129--140

\bibitem[{{Brott} \& {Hauschildt}(2005)}]{2005ESASP.576..565B}
{Brott}, I. \& {Hauschildt}, P.~H. 2005, in ESA Special Publication, Vol. 576,
  The Three-Dimensional Universe with Gaia, ed. C.~{Turon}, K.~S. {O'Flaherty},
  \& M.~A.~C. {Perryman}, 565

\bibitem[{{Brown}(2015)}]{2015ApJ...799...87B}
{Brown}, R.~A. 2015, \apj, 799, 87

\bibitem[{{Defr{\`e}re} {et~al.}(2010){Defr{\`e}re}, {Absil}, {den Hartog},
  {Hanot}, \& {Stark}}]{2010A&A...509A...9D}
{Defr{\`e}re}, D., {Absil}, O., {den Hartog}, R., {Hanot}, C., \& {Stark}, C.
  2010, \aap, 509, A9

\bibitem[{{Defr{\`e}re} {et~al.}(2015){Defr{\`e}re}, {Hinz}, {Skemer},
  {Kennedy}, {Bailey}, {Hoffmann}, {Mennesson}, {Millan-Gabet}, {Danchi},
  {Absil}, {Arbo}, {Beichman}, {Brusa}, {Bryden}, {Downey}, {Durney},
  {Esposito}, {Gaspar}, {Grenz}, {Haniff}, {Hill}, {Lebreton}, {Leisenring},
  {Males}, {Marion}, {McMahon}, {Montoya}, {Morzinski}, {Pinna}, {Puglisi},
  {Rieke}, {Roberge}, {Serabyn}, {Sosa}, {Stapeldfeldt}, {Su}, {Vaitheeswaran},
  {Vaz}, {Weinberger}, \& {Wyatt}}]{2015ApJ...799...42D}
{Defr{\`e}re}, D., {Hinz}, P.~M., {Skemer}, A.~J., {Kennedy}, G.~M., {Bailey},
  V.~P., {Hoffmann}, W.~F., {Mennesson}, B., {Millan-Gabet}, R., {Danchi},
  W.~C., {Absil}, O., {Arbo}, P., {Beichman}, C., {Brusa}, G., {Bryden}, G.,
  {Downey}, E.~C., {Durney}, O., {Esposito}, S., {Gaspar}, A., {Grenz}, P.,
  {Haniff}, C., {Hill}, J.~M., {Lebreton}, J., {Leisenring}, J.~M., {Males},
  J.~R., {Marion}, L., {McMahon}, T.~J., {Montoya}, M., {Morzinski}, K.~M.,
  {Pinna}, E., {Puglisi}, A., {Rieke}, G., {Roberge}, A., {Serabyn}, E.,
  {Sosa}, R., {Stapeldfeldt}, K., {Su}, K., {Vaitheeswaran}, V., {Vaz}, A.,
  {Weinberger}, A.~J., \& {Wyatt}, M.~C. 2015, \apj, 799, 42

\bibitem[{{Dermott} {et~al.}(2002){Dermott}, {Durda}, {Grogan}, \&
  {Kehoe}}]{2002aste.conf..423D}
{Dermott}, S.~F., {Durda}, D.~D., {Grogan}, K., \& {Kehoe}, T.~J.~J. 2002,
  Asteroids III, 423

\bibitem[{{Eiroa} {et~al.}(2013)}]{2013A&A...555A..11E}
{Eiroa}, C. {et~al.} 2013, \aap, 555, A11

\bibitem[{{Fujiwara} {et~al.}(2010){Fujiwara}, {Onaka}, {Ishihara},
  {Yamashita}, {Fukagawa}, {Nakagawa}, {Kataza}, {Ootsubo}, \&
  {Murakami}}]{2010ApJ...714L.152F}
{Fujiwara}, H., {Onaka}, T., {Ishihara}, D., {Yamashita}, T., {Fukagawa}, M.,
  {Nakagawa}, T., {Kataza}, H., {Ootsubo}, T., \& {Murakami}, H. 2010, \apjl,
  714, L152

\bibitem[{{G{\'a}sp{\'a}r} {et~al.}(2013){G{\'a}sp{\'a}r}, {Rieke}, \&
  {Balog}}]{2013ApJ...768...25G}
{G{\'a}sp{\'a}r}, A., {Rieke}, G.~H., \& {Balog}, Z. 2013, \apj, 768, 25

\bibitem[{{Grogan} {et~al.}(2001){Grogan}, {Dermott}, \&
  {Durda}}]{2001Icar..152..251G}
{Grogan}, K., {Dermott}, S.~F., \& {Durda}, D.~D. 2001, \icarus, 152, 251

\bibitem[{{Hinz}(2009)}]{2009AIPC.1158..313H}
{Hinz}, P.~M. 2009, in American Institute of Physics Conference Series, Vol.
  1158, American Institute of Physics Conference Series, ed. T.~{Usuda},
  M.~{Tamura}, \& M.~{Ishii}, 313--317

\bibitem[{{Johnson} {et~al.}(2008){Johnson}, {Marcy}, {Fischer}, {Wright},
  {Reffert}, {Kregenow}, {Williams}, \& {Peek}}]{2008ApJ...675..784J}
{Johnson}, J.~A., {Marcy}, G.~W., {Fischer}, D.~A., {Wright}, J.~T., {Reffert},
  S., {Kregenow}, J.~M., {Williams}, P.~K.~G., \& {Peek}, K.~M.~G. 2008, \apj,
  675, 784

\bibitem[{{Kains} {et~al.}(2011){Kains}, {Wyatt}, \&
  {Greaves}}]{2011MNRAS.414.2486K}
{Kains}, N., {Wyatt}, M.~C., \& {Greaves}, J.~S. 2011, \mnras, 414, 2486

\bibitem[{{Kelsall} {et~al.}(1998){Kelsall}, {Weiland}, {Franz}, {Reach},
  {Arendt}, {Dwek}, {Freudenreich}, {Hauser}, {Moseley}, {Odegard},
  {Silverberg}, \& {Wright}}]{1998ApJ...508...44K}
{Kelsall}, T., {Weiland}, J.~L., {Franz}, B.~A., {Reach}, W.~T., {Arendt},
  R.~G., {Dwek}, E., {Freudenreich}, H.~T., {Hauser}, M.~G., {Moseley}, S.~H.,
  {Odegard}, N.~P., {Silverberg}, R.~F., \& {Wright}, E.~L. 1998, \apj, 508, 44

\bibitem[{{Kennedy} \& {Wyatt}(2013)}]{2013MNRAS.433.2334K}
{Kennedy}, G.~M. \& {Wyatt}, M.~C. 2013, \mnras, 433, 2334

\bibitem[{{Kennedy} {et~al.}(2015){Kennedy}, {Wyatt}, {Bailey}, {Bryden},
  {Danchi}, {Defr{\`e}re}, {Haniff}, {Hinz}, {Lebreton}, {Mennesson},
  {Millan-Gabet}, {Morales}, {Pani{\'c}}, {Rieke}, {Roberge}, {Serabyn},
  {Shannon}, {Skemer}, {Stapelfeldt}, {Su}, \&
  {Weinberger}}]{2015ApJS..216...23K}
{Kennedy}, G.~M., {Wyatt}, M.~C., {Bailey}, V., {Bryden}, G., {Danchi}, W.~C.,
  {Defr{\`e}re}, D., {Haniff}, C., {Hinz}, P.~M., {Lebreton}, J., {Mennesson},
  B., {Millan-Gabet}, R., {Morales}, F., {Pani{\'c}}, O., {Rieke}, G.~H.,
  {Roberge}, A., {Serabyn}, E., {Shannon}, A., {Skemer}, A.~J., {Stapelfeldt},
  K.~R., {Su}, K.~Y.~L., \& {Weinberger}, A.~J. 2015, \apjs, 216, 23

\bibitem[{{Kennedy} {et~al.}(2012{\natexlab{a}}){Kennedy}, {Wyatt},
  {Sibthorpe}, {Phillips}, {Matthews}, \& {Greaves}}]{2012MNRAS.426.2115K}
{Kennedy}, G.~M., {Wyatt}, M.~C., {Sibthorpe}, B., {Phillips}, N.~M.,
  {Matthews}, B.~C., \& {Greaves}, J.~S. 2012{\natexlab{a}}, \mnras, 426, 2115

\bibitem[{{Kennedy} {et~al.}(2012{\natexlab{b}})}]{2012MNRAS.421.2264K}
{Kennedy}, G.~M. {et~al.} 2012{\natexlab{b}}, \mnras, 421, 2264

\bibitem[{{Krijt} \& {Kama}(2014)}]{2014A&A...566L...2K}
{Krijt}, S. \& {Kama}, M. 2014, \aap, 566, L2

\bibitem[{{Kuchner} \& {Stark}(2010)}]{2010AJ....140.1007K}
{Kuchner}, M.~J. \& {Stark}, C.~C. 2010, \aj, 140, 1007

\bibitem[{{Lay}(2004)}]{2004ApOpt..43.6100L}
{Lay}, O.~P. 2004, \ao, 43, 6100

\bibitem[{{Lestrade} {et~al.}(2012)}]{2012A&A...548A..86L}
{Lestrade}, J.-F. {et~al.} 2012, \aap, 548, A86

\bibitem[{{Liou} \& {Zook}(1999)}]{1999AJ....118..580L}
{Liou}, J.-C. \& {Zook}, H.~A. 1999, \aj, 118, 580

\bibitem[{{Liou} {et~al.}(1996){Liou}, {Zook}, \&
  {Dermott}}]{1996Icar..124..429L}
{Liou}, J.-C., {Zook}, H.~A., \& {Dermott}, S.~F. 1996, \icarus, 124, 429

\bibitem[{{Lisse} {et~al.}(2012){Lisse}, {Wyatt}, {Chen}, {Morlok}, {Watson},
  {Manoj}, {Sheehan}, {Currie}, {Thebault}, \& {Sitko}}]{2012ApJ...747...93L}
{Lisse}, C.~M., {Wyatt}, M.~C., {Chen}, C.~H., {Morlok}, A., {Watson}, D.~M.,
  {Manoj}, P., {Sheehan}, P., {Currie}, T.~M., {Thebault}, P., \& {Sitko},
  M.~L. 2012, \apj, 747, 93

\bibitem[{{Meng} {et~al.}(2012){Meng}, {Rieke}, {Su}, {Ivanov}, {Vanzi}, \&
  {Rujopakarn}}]{2012ApJ...751L..17M}
{Meng}, H.~Y.~A., {Rieke}, G.~H., {Su}, K.~Y.~L., {Ivanov}, V.~D., {Vanzi}, L.,
  \& {Rujopakarn}, W. 2012, \apjl, 751, L17

\bibitem[{{Mennesson} {et~al.}(2014){Mennesson}, {Millan-Gabet}, {Serabyn},
  {Colavita}, {Absil}, {Bryden}, {Wyatt}, {Danchi}, {Defr{\`e}re}, {Dor{\'e}},
  {Hinz}, {Kuchner}, {Ragland}, {Scott}, {Stapelfeldt}, {Traub}, \&
  {Woillez}}]{2014ApJ...797..119M}
{Mennesson}, B., {Millan-Gabet}, R., {Serabyn}, E., {Colavita}, M.~M., {Absil},
  O., {Bryden}, G., {Wyatt}, M., {Danchi}, W., {Defr{\`e}re}, D., {Dor{\'e}},
  O., {Hinz}, P., {Kuchner}, M., {Ragland}, S., {Scott}, N., {Stapelfeldt}, K.,
  {Traub}, W., \& {Woillez}, J. 2014, \apj, 797, 119

\bibitem[{{Millan-Gabet} {et~al.}(2011){Millan-Gabet}, {Serabyn}, {Mennesson},
  {Traub}, {Barry}, {Danchi}, {Kuchner}, {Stark}, {Ragland}, {Hrynevych},
  {Woillez}, {Stapelfeldt}, {Bryden}, {Colavita}, \&
  {Booth}}]{2011ApJ...734...67M}
{Millan-Gabet}, R., {Serabyn}, E., {Mennesson}, B., {Traub}, W.~A., {Barry},
  R.~K., {Danchi}, W.~C., {Kuchner}, M., {Stark}, C.~C., {Ragland}, S.,
  {Hrynevych}, M., {Woillez}, J., {Stapelfeldt}, K., {Bryden}, G., {Colavita},
  M.~M., \& {Booth}, A.~J. 2011, \apj, 734, 67

\bibitem[{{Nesvorn{\'y}} {et~al.}(2010){Nesvorn{\'y}}, {Jenniskens}, {Levison},
  {Bottke}, {Vokrouhlick{\'y}}, \& {Gounelle}}]{2010ApJ...713..816N}
{Nesvorn{\'y}}, D., {Jenniskens}, P., {Levison}, H.~F., {Bottke}, W.~F.,
  {Vokrouhlick{\'y}}, D., \& {Gounelle}, M. 2010, \apj, 713, 816

\bibitem[{{Pawellek} {et~al.}(2014){Pawellek}, {Krivov}, {Marshall},
  {Montesinos}, {{\'A}brah{\'a}m}, {Mo{\'o}r}, {Bryden}, \&
  {Eiroa}}]{2014ApJ...792...65P}
{Pawellek}, N., {Krivov}, A.~V., {Marshall}, J.~P., {Montesinos}, B.,
  {{\'A}brah{\'a}m}, P., {Mo{\'o}r}, A., {Bryden}, G., \& {Eiroa}, C. 2014,
  \apj, 792, 65

\bibitem[{{Phillips} {et~al.}(2010){Phillips}, {Greaves}, {Dent}, {Matthews},
  {Holland}, {Wyatt}, \& {Sibthorpe}}]{2010MNRAS.403.1089P}
{Phillips}, N.~M., {Greaves}, J.~S., {Dent}, W.~R.~F., {Matthews}, B.~C.,
  {Holland}, W.~S., {Wyatt}, M.~C., \& {Sibthorpe}, B. 2010, \mnras, 403, 1089

\bibitem[{{Rauer} {et~al.}(2014)}]{2014ExA....38..249R}
{Rauer}, H. {et~al.} 2014, Experimental Astronomy, 38, 249

\bibitem[{{Reidemeister} {et~al.}(2011){Reidemeister}, {Krivov}, {Stark},
  {Augereau}, {L{\"o}hne}, \& {M{\"u}ller}}]{2011A&A...527A..57R}
{Reidemeister}, M., {Krivov}, A.~V., {Stark}, C.~C., {Augereau}, J.-C.,
  {L{\"o}hne}, T., \& {M{\"u}ller}, S. 2011, \aap, 527, A57

\bibitem[{{Rieke} {et~al.}(2005)}]{2005ApJ...620.1010R}
{Rieke}, G.~H. {et~al.} 2005, \apj, 620, 1010

\bibitem[{{Roberge} {et~al.}(2012)}]{2012PASP..124..799R}
{Roberge}, A. {et~al.} 2012, \pasp, 124, 799

\bibitem[{{Rodriguez} \& {Zuckerman}(2012)}]{2012ApJ...745..147R}
{Rodriguez}, D.~R. \& {Zuckerman}, B. 2012, \apj, 745, 147

\bibitem[{{Sch{\"u}ppler} {et~al.}(2014){Sch{\"u}ppler}, {L{\"o}hne}, {Krivov},
  {Ertel}, {Marshall}, \& {Eiroa}}]{2014A&A...567A.127S}
{Sch{\"u}ppler}, C., {L{\"o}hne}, T., {Krivov}, A.~V., {Ertel}, S., {Marshall},
  J.~P., \& {Eiroa}, C. 2014, \aap, 567, A127

\bibitem[{{Serabyn} {et~al.}(2012){Serabyn}, {Mennesson}, {Colavita},
  {Koresko}, \& {Kuchner}}]{2012ApJ...748...55S}
{Serabyn}, E., {Mennesson}, B., {Colavita}, M.~M., {Koresko}, C., \& {Kuchner},
  M.~J. 2012, \apj, 748, 55

\bibitem[{{Sierchio} {et~al.}(2014){Sierchio}, {Rieke}, {Su}, \&
  {G{\'a}sp{\'a}r}}]{2014ApJ...785...33S}
{Sierchio}, J.~M., {Rieke}, G.~H., {Su}, K.~Y.~L., \& {G{\'a}sp{\'a}r}, A.
  2014, \apj, 785, 33

\bibitem[{{Smith} {et~al.}(2009){Smith}, {Wyatt}, \&
  {Haniff}}]{2009A&A...503..265S}
{Smith}, R., {Wyatt}, M.~C., \& {Haniff}, C.~A. 2009, \aap, 503, 265

\bibitem[{{Song} {et~al.}(2005){Song}, {Zuckerman}, {Weinberger}, \&
  {Becklin}}]{2005Natur.436..363S}
{Song}, I., {Zuckerman}, B., {Weinberger}, A.~J., \& {Becklin}, E.~E. 2005,
  \nat, 436, 363

\bibitem[{{Stark} \& {Kuchner}(2008)}]{2008ApJ...686..637S}
{Stark}, C.~C. \& {Kuchner}, M.~J. 2008, \apj, 686, 637

\bibitem[{{Stark} {et~al.}(2014){Stark}, {Roberge}, {Mandell}, \&
  {Robinson}}]{2014ApJ...795..122S}
{Stark}, C.~C., {Roberge}, A., {Mandell}, A., \& {Robinson}, T.~D. 2014, \apj,
  795, 122

\bibitem[{{Su} {et~al.}(2006){Su}, {Rieke}, {Stansberry}, {Bryden},
  {Stapelfeldt}, {Trilling}, {Muzerolle}, {Beichman}, {Moro-Martin}, {Hines},
  \& {Werner}}]{2006ApJ...653..675S}
{Su}, K.~Y.~L., {Rieke}, G.~H., {Stansberry}, J.~A., {Bryden}, G.,
  {Stapelfeldt}, K.~R., {Trilling}, D.~E., {Muzerolle}, J., {Beichman}, C.~A.,
  {Moro-Martin}, A., {Hines}, D.~C., \& {Werner}, M.~W. 2006, \apj, 653, 675

\bibitem[{{Thureau} {et~al.}(2014){Thureau}, {Greaves}, {Matthews}, {Kennedy},
  {Phillips}, {Booth}, {Duch{\^e}ne}, {Horner}, {Rodriguez}, {Sibthorpe}, \&
  {Wyatt}}]{2014MNRAS.445.2558T}
{Thureau}, N.~D., {Greaves}, J.~S., {Matthews}, B.~C., {Kennedy}, G.,
  {Phillips}, N., {Booth}, M., {Duch{\^e}ne}, G., {Horner}, J., {Rodriguez},
  D.~R., {Sibthorpe}, B., \& {Wyatt}, M.~C. 2014, \mnras, 445, 2558

\bibitem[{{Trilling} {et~al.}(2008){Trilling}, {Bryden}, {Beichman}, {Rieke},
  {Su}, {Stansberry}, {Blaylock}, {Stapelfeldt}, {Beeman}, \&
  {Haller}}]{2008ApJ...674.1086T}
{Trilling}, D.~E., {Bryden}, G., {Beichman}, C.~A., {Rieke}, G.~H., {Su},
  K.~Y.~L., {Stansberry}, J.~A., {Blaylock}, M., {Stapelfeldt}, K.~R.,
  {Beeman}, J.~W., \& {Haller}, E.~E. 2008, \apj, 674, 1086

\bibitem[{{van Lieshout} {et~al.}(2014){van Lieshout}, {Dominik}, {Kama}, \&
  {Min}}]{2014A&A...571A..51V}
{van Lieshout}, R., {Dominik}, C., {Kama}, M., \& {Min}, M. 2014, \aap, 571,
  A51

\bibitem[{{Vitense} {et~al.}(2012){Vitense}, {Krivov}, {Kobayashi}, \&
  {L{\"o}hne}}]{2012A&A...540A..30V}
{Vitense}, C., {Krivov}, A.~V., {Kobayashi}, H., \& {L{\"o}hne}, T. 2012, \aap,
  540, A30

\bibitem[{{Weinberger} {et~al.}(2015){Weinberger}, {Bryden}, {Kennedy},
  {Roberge}, {Defr{\`e}re}, {Hinz}, {Millan-Gabet}, {Rieke}, {Bailey},
  {Danchi}, {Haniff}, {Mennesson}, {Serabyn}, {Skemer}, {Stapelfeldt}, \&
  {Wyatt}}]{2015ApJS..216...24W}
{Weinberger}, A.~J., {Bryden}, G., {Kennedy}, G.~M., {Roberge}, A.,
  {Defr{\`e}re}, D., {Hinz}, P.~M., {Millan-Gabet}, R., {Rieke}, G., {Bailey},
  V.~P., {Danchi}, W.~C., {Haniff}, C., {Mennesson}, B., {Serabyn}, E.,
  {Skemer}, A.~J., {Stapelfeldt}, K.~R., \& {Wyatt}, M.~C. 2015, \apjs, 216, 24

\bibitem[{{Whipple} {et~al.}(1967){Whipple}, {Southworth}, \&
  {Nilsson}}]{1967SAOSR.239.....W}
{Whipple}, F.~L., {Southworth}, R.~B., \& {Nilsson}, C.~S. 1967, SAO Special
  Report, 239

\bibitem[{{Wyatt}(2005)}]{2005A&A...433.1007W}
{Wyatt}, M.~C. 2005, \aap, 433, 1007

\bibitem[{{Wyatt}(2008)}]{2008ARA&A..46..339W}
---. 2008, \araa, 46, 339

\bibitem[{{Wyatt} {et~al.}(1999){Wyatt}, {Dermott}, {Telesco}, {Fisher},
  {Grogan}, {Holmes}, \& {Pi{\~n}a}}]{1999ApJ...527..918W}
{Wyatt}, M.~C., {Dermott}, S.~F., {Telesco}, C.~M., {Fisher}, R.~S., {Grogan},
  K., {Holmes}, E.~K., \& {Pi{\~n}a}, R.~K. 1999, \apj, 527, 918

\bibitem[{{Wyatt} {et~al.}(2005){Wyatt}, {Greaves}, {Dent}, \&
  {Coulson}}]{2005ApJ...620..492W}
{Wyatt}, M.~C., {Greaves}, J.~S., {Dent}, W.~R.~F., \& {Coulson}, I.~M. 2005,
  \apj, 620, 492

\bibitem[{{Wyatt} {et~al.}(2007{\natexlab{a}}){Wyatt}, {Smith}, {Greaves},
  {Beichman}, {Bryden}, \& {Lisse}}]{2007ApJ...658..569W}
{Wyatt}, M.~C., {Smith}, R., {Greaves}, J.~S., {Beichman}, C.~A., {Bryden}, G.,
  \& {Lisse}, C.~M. 2007{\natexlab{a}}, \apj, 658, 569

\bibitem[{{Wyatt} {et~al.}(2007{\natexlab{b}}){Wyatt}, {Smith}, {Su}, {Rieke},
  {Greaves}, {Beichman}, \& {Bryden}}]{2007ApJ...663..365W}
{Wyatt}, M.~C., {Smith}, R., {Su}, K.~Y.~L., {Rieke}, G.~H., {Greaves}, J.~S.,
  {Beichman}, C.~A., \& {Bryden}, G. 2007{\natexlab{b}}, \apj, 663, 365

\end{thebibliography}

\end{document}